\documentclass[a4paper,prl,twocolumn,showpacs,10pt,aps,reprint,superscriptaddress]{revtex4-1}

\usepackage{lmodern} 

\usepackage{amsmath,amssymb,hyperref}
\usepackage{graphicx}
\usepackage{datetime}
\usepackage{xfrac}
\usepackage[usenames,dvipsnames]{color}

\definecolor{MyGreen}{rgb}{0,0.6,0}

\newcommand{\figref}[1]{Fig.~\ref{#1}}
\newcommand{\eqnref}[1]{Eqn.~(\ref{#1})}

\newcommand{\micron}{\ensuremath{\mu{\rm m}}}
\newcommand{\partialderiv}[2]{\frac{\partial #1}{\partial #2}}

\begin{document}


\title{Bose-Einstein condensation of photons from the thermodynamic limit to small photon numbers}


\author{Robert A. Nyman}\email[Correspondence to
]{r.nyman@imperial.ac.uk} 
\affiliation{Quantum Optics and Laser Science group,Blackett Laboratory, 
Imperial College London, Prince Consort Road, SW7 2BW, United Kingdom}
\author{Benjamin T. Walker}
\affiliation{Quantum Optics and Laser Science group,Blackett Laboratory, 
Imperial College London, Prince Consort Road, SW7 2BW, United Kingdom}
\affiliation{Centre for Doctoral Training in Cont	olled Quantum Dynamics,
Imperial College London, Prince Consort Road, SW7 2BW, United Kingdom}

\begin{abstract}
  Photons can come to thermal equilibrium at room temperature by scattering multiple times from a fluorescent dye. By confining the light and dye in a microcavity, a minimum energy is set and the photons can then show Bose-Einstein condensation. We present here the physical principles underlying photon thermalization and condensation, and review the literature on the subject. We then explore the `small' regime where very few photons are needed for condensation. We compare thermal equilibrium results to a rate-equation model of microlasers, which includes spontaneous emission into the cavity, and we note that small systems result in ambiguity in the definition of threshold.
\end{abstract}

\maketitle

\section{Foreword}

This article is written in memory of Danny Segal, who was a colleague of one of us (Rob Nyman) in the Quantum Optics and Laser Science group at Imperial College for many years. The topic of this article touches on the subject of dye lasers, the stuff of nightmares for any AMO physicist of his generation, but a stronger connection to Danny is that he was very supportive of my application for the fellowship that pushed my career forward, and funded this research. One of Danny's quirks was a strong dislike of flying. As a consequence, I had the pleasure of joining him on a 24~hour, four-train journey from London to Italy to a conference. That's a lot of time for story telling and forging memories for life. Danny was one of the good guys, and I sorely miss his good humour and advice. 

This article presents a gentle introduction to thermalization and Bose-Einstein condensation (BEC) of photons in dye-filled microcavities, followed by a review of the state of the art. We then note the similarity to microlasers, particularly when there are very few photons involved. We compare a simple non-equilibrium model for microlasers with an even simpler thermal equilibrium model for BEC and show that the models coincide for similar values of a `smallness' parameter.

\section{Introduction to thermalization and condensation of photons}


Bose-Einstein condensation can be defined as macroscopic occupation of the ground state at thermal equilibrium~\cite{Pathria}\footnote{The formal Penrose-Onsager definition~\cite{Penrose56} is essentially that the largest eigenvalue of the single-particle density matrix that solves the full many-body problem is extensive with system size. But that's not the kind of definition that helps the non-expert.}. It is a natural consequence of the exchange statistics of identical bosons, and therefore occurs in a wide variety of physical systems, such as liquid helium, ultracold atomic gases or electron pairs in superconductors.

The Bose-Einstein distribution for non-interacting identical bosons may be familiar to most physicists, but it's worth looking at again: 
\begin{equation}
  f(\epsilon) = {1}/\left[{\exp{(\epsilon-\mu)/k_B T}-1}\right]
  , \label{eqn:BE distribution}
\end{equation}
which gives the occupancy $f$ of a state at energy $\epsilon$. It has two parameters. $T$ is the temperature, which tells us immediately that we are discussing thermal-equilibrium phenomena. The chemical potential is $\mu$, which is the thermal energy required to add another particle to the system from a reservoir, and it dictates the number of particles in the system. The chemical potential is always lower than the energy of the ground state in the system, but as it approaches from below, the distribution shows a divergence in the ground-state population. That's BEC.

Photons can be brought to thermal equilibrium in a black box, but their number is not conserved, which implies that the chemical potential is either not well defined or strictly zero, depending on your point of view. In either case, BEC is not possible. It is however possible to give light a non-zero chemical potential in a medium which has an optical transition between two broad bands, such as a semiconductor (with valence and conduction bands) or a fluorescent dye (with ro-vibrationally broadened electronic states), as explained by W\"urfel~\cite{Wurfel82}. The chemical potential then sets the population of the excited band. There is a kind of chemical equilibrium between photons and excitations in the medium (induced for example by optical pumping), with reactions being absorption and emission. Thus the chemical potential of the photons will equal that of the medium, which is non-zero, provided there are enough absorption and emission events, i.e. enough time for the chemical reaction to have taken place before other loss processes occur or the photons leave the system.

\begin{figure}[htb]
	\centering
	\includegraphics[width=0.95\columnwidth]{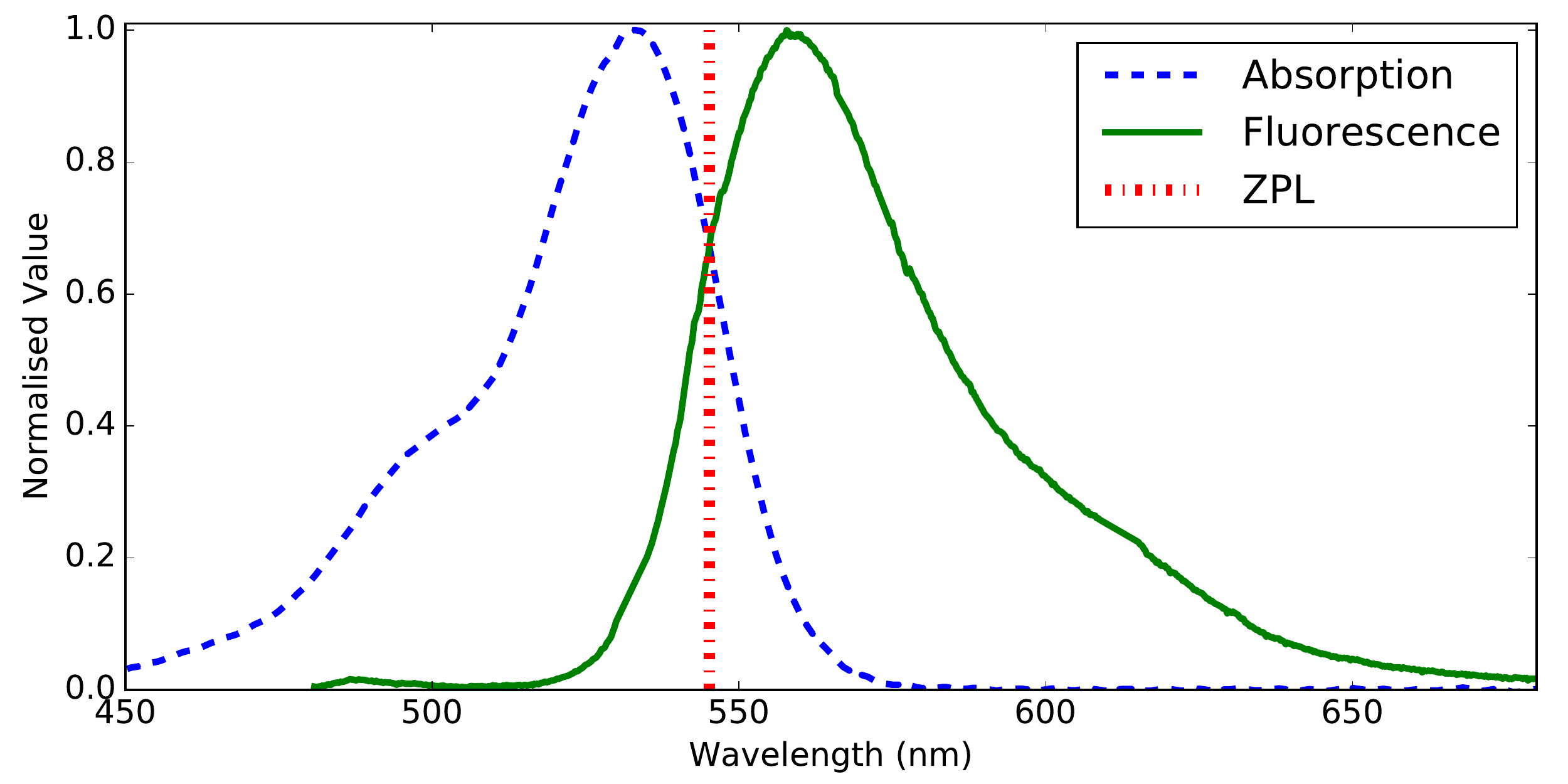}
	\caption{
	Absorption and fluorescence emission spectra for Rhodamine~6G in Ethylene Glycol, the solution most used for thermalizing and condensing photons. Both are normalized to their peak value, and the ZPL is the wavelength at which they are equal. 
	The data in this figure are available from Ref.~\cite{Nyman17Rhodamine}.
	}
	\label{fig:abs and fluo}
\end{figure}

Not only will the photons acquire a non-zero chemical potential, they will also reach a thermal equilibrium population, dictated by the ratio of absorption (loss) and emission (gain) of the optically interesting medium. That relation is known as the Kennard-Stepanov~\cite{Kennard26}\footnote{The canonical citation of Stepanov is Ref.~\cite{Stepanov57}, but we cannot find a copy of it, nor read Russian.} or McCumber relation~\cite{McCumber64}. The ratio is dictated by a principle of detailed balance, through rapid relaxation among the states within the bands, i.e. vibrational relaxation of dye molecules, which takes perhaps 100~fs, which is fast compared to typical spontaneous emission lifetimes of a few ns. It is usually possible to identify a zero-phonon line (ZPL) about which the spectra are symmetric with a Boltzmann factor between them:

\begin{equation}
 \frac{A(\epsilon)}{F(\epsilon)} =  \rm{e}^{(\epsilon-\epsilon_{ZPL})/k_B T} \label{eqn:KSR}
\end{equation}
where $F$ and $A$ are fluorescence emission and absorption respectively, normalized to their peak values, $\epsilon$ is the energy of the light and $\epsilon_{ZPL}$ the energy of the ZPL~\footnote{For the sake of simplicity we have omitted a factor $\epsilon^3$ which accounts for the density of states available for spontaneous emission.}. The temperature $T$ is the temperature of the the phonons that cause vibrational relaxation, assumed to be the same as the temperature of the bulk medium. In \figref{fig:abs and fluo} we show the absorption and emission spectra of Rhodamine~6G in ethylene glycol, which shows the expected symmetry. Analysis reveals that the Kennard-Stepanov/McCumber relation \eqnref{eqn:KSR} is very well matched at room temperature over a wide range of wavelengths.

But BEC requires a unique ground state into which condensation can occur. We engineer the density of states for the light using a pair of near-planar mirrors, a Fabry-Perot optical cavity. Let us make the mirrors so close together that the free spectral range, the difference between longitudinal modes of the cavity, is at least as large as the widths of the absorption and emission spectra of the dye. Then, only one longitudinal mode is relevant, but there can be many transverse modes. This is known as a microcavity. For photon thermalization in Rhodamine, the cavity is typically about 8 half wavelengths long, so we label the longitudinal mode $q=8$. 

The cavity-resonant energy is minimum for light propagating parallel to the cavity optical axis. Light propagating at an angle must have higher energy to match the boundary conditions for resonance. For small angles, the energy is quadratically proportional to the in-plane wavenumber (or momentum) of the light, just like a massive particle with with kinetic energy. 

We can understand how the transverse modes of the cavity relate to the shape of the mirrors, by considering a local cavity length, and hence local cavity-resonant energy for the light. Where the mirrors are closer together, the energy is higher, so there is an effective potential energy cost, dependent on the mirror shape. 

Thus the light can be considered as a massive particle moving in a trapping potential (assuming that the cavity is convex, longer in the middle than the edges), whose energy as a function of momentum ${\bf p}$ and position ${\bf r}$ is:
\begin{align}
  E({\bf r, p}) = m {c^*}^2 + \frac{p^2}{2 m} + V({\bf r})
  \label{eqn:particle energy}
\end{align}
Here the effective mass is given by the cavity length $L_0$ and $c^*=c/n$ the speed of light in the intracavity medium: $m = h n^2 / c\lambda_0$. The cutoff wavelength $\lambda_0$ is the longest wavelength (for light emitted from the cavity) which is resonant with the cavity in the pertinent longitudinal mode: $\lambda_0 = 2n L_0 / q$. The local potential energy $V({\bf r})$ is given by local deviations of the cavity length, $\delta L({\bf r})$ and can be simply written as $V({\bf r}) / m {c^*}^2 = \delta L({\bf r})/L_0$.

\subsection{Experimental apparatus}

\begin{figure}[htb]
	\centering
	\includegraphics[width=0.95\columnwidth]{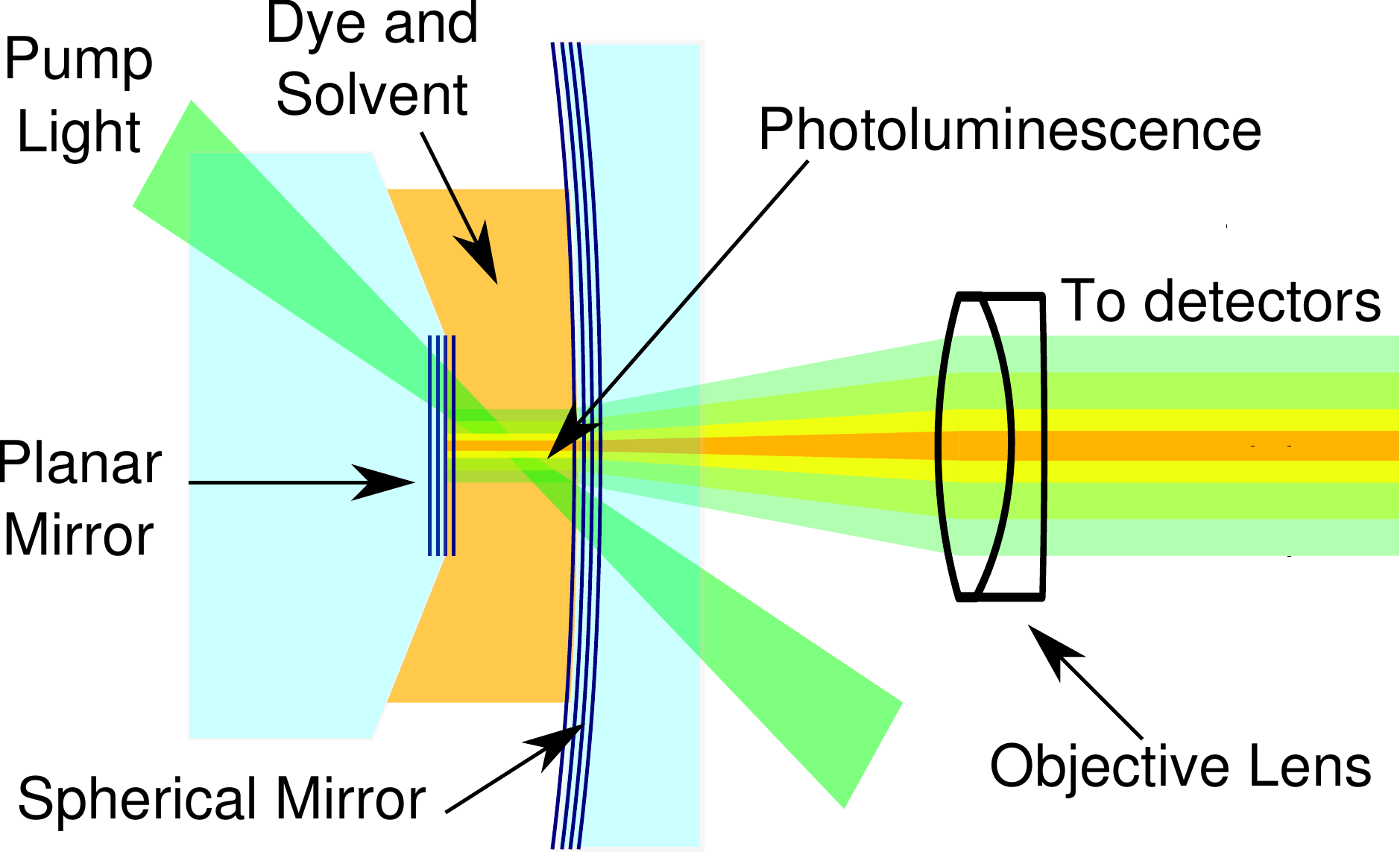}
	\includegraphics[width=0.6\columnwidth]{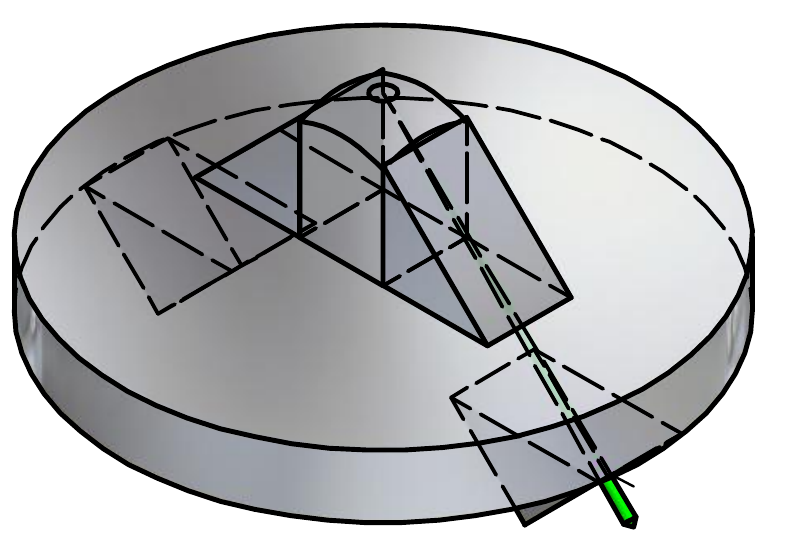}
	\caption{
	Top: Schematic of the apparatus required for photon thermalization and condensation. Because the distance between the mirrors is short (about 1.5~\micron) and one mirror is curved the planar mirror is ground down to about 1~mm diameter. To pump at an angle, taking advantage of an angle-dependent transparency of the dielectric mirrors, the planar mirror is built into a simple optical assembly (bottom).
	}
	\label{fig:apparatus}
\end{figure}

To trap the light long enough for thermalization through multiple absorption events, the cavity mirrors must have reflectivity of at least 99.99\% ($<100$~ppm loss). Such mirrors are commercially available, and use dielectric coatings of several pairs of layers. The simplest configuration for a cavity uses spherical mirrors, either a pair, or one in conjunction with a planar mirror, as shown in \figref{fig:apparatus}. The spherical cavity length variation leads to a harmonic potential, at least close to the longest part of the cavity, where the photons are trapped. Typically the mirror radius of curvature is about 0.5 m, leading to mode spacings (trapping frequencies) around 40~GHz. Because of the curvature and the proximity of the mirrors, being just a few half-wavelengths apart, at least one of mirrors is ground down to about 1~mm diameter. Light is pumped at an angle to the optic axis, to take advantage of the transmission maximum of the dielectric mirror coating, As a result, the mirrors are often glued to other components which make alignment of the pump easier: see \figref{fig:apparatus} for one example of how the assembly can be done.

To align the cavity, the mirrors require five degrees of freedom for their relative position and orientation. The separation on the optic axis must be actively controlled with nanometre-precision using for example a piezo-electric translation stage. It is very likely that the cavity length will need to be actively stabilized, yet scanned so that the resonant wavelength varies by tens of nanometers. The solution is to shine a collimated beam of narrowband light at the edge of the stop-band of the mirrors, at which wavelength they transmit at least an order of magnitude more than at the wavelengths used for thermalized light, and the dye does not strongly absorb this light. HeNe laser light 633~nm wavelength is a good match to mirrors designed for thermalizing light at around 580~nm using Rhodamine~6G. This narrowband light then forms rings, similar to Newton's rings. The images are acquired by a camera, processed to find the ring radius, and feedback is applied to actively control the cavity length.

Light emitted from the cavity is collected by an objective. Our imaging system uses a simple achromatic doublet in an afocal setup, i.e. producing an image at infinity. This collimated light is then split, by dichroic mirrors (to extract the stabilization reference light) and by non-polarizing beamsplitters, after which is it sent to a variety of diagnostic tools. The most important tools are a camera and a spectrometer. 

\begin{figure}[htb]
	\centering
	\includegraphics[width=0.57\columnwidth]{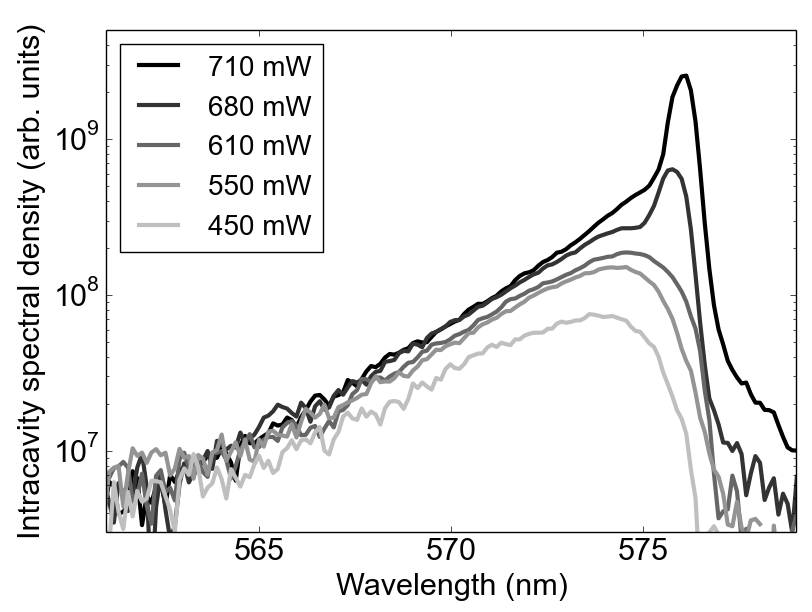}
	\includegraphics[width=0.41\columnwidth]{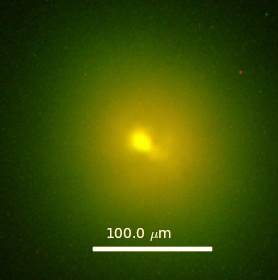}	
	\caption{
	Typical data showing photon BEC. Left: spectra. At low pump powers, the spectrum is compatible with the Boltzmann distribution at room temperature, showing both a cutoff (ground state) and a density of states equivalent to a two-dimensional harmonic oscillator. At higher powers, extra photons go into the ground state and the populations of the excited states saturate. Right: a real-space, real-colour image just above threshold, showing a thermal cloud of photons around a large population in the centre, where the lowest-energy transverse mode of the cavity is located.
	}
	\label{fig:characteristic results}
\end{figure}

Typical data are shown in \figref{fig:characteristic results}. The spectrum shows, at low power, a distribution which is compatible with the Bose-Einstein distribution, \eqnref{eqn:BE distribution}, at room temperature, taking into account the density of states for a two-dimensional (2D) symmetric harmonic trapping potential. There is a clear cutoff showing that there is a well-defined lowest-energy mode, in this case around $\lambda_0 = 576$~nm. As the pump power is increased, the chemical potential approaches zero from below, and the population of the lowest mode increases dramatically while all other modes saturate. BEC is the explanation. The image is taken just above threshold. It shows a Gaussian fuzz around, which is the non-condensed, thermal cloud, whose size depends mainly on the temperature and the trapping potential. In the centre is a bright spot, indicating the large occupation of the smallest mode, which is the lowest energy. It is noteworthy that the light is green (high energy) at the edge, where the potential energy is highest, and yellow (low energy) in the middle. At high pump intensities, the pump must be pulsed with a low duty cycle, so that the population of a scientifically-uninteresting triplet state~\cite{Schaefer, Coles15} is kept low. Typically, pulses last about 500~ns and are repeated at about 500~Hz.

\subsection{Things which are like Photon BEC but which are not Photon BEC}


At this point we will digress and discuss three other kinds of condensates of light, none of which is photon BEC: exciton-polariton condensates; classical wave condensation of light and BEC of plasmons.

There is a large community working with light and solid-state matter which are strongly-coupled, in the cavity QED sense that the coherent coupling is faster than incoherent mechanisms like spontaneous emission or cavity loss, using microcavities. Strongly coupled light-matter systems are known as polaritons. Typically the light interacts with a quasiparticle made of a bound electron and hole pair known as an exciton, making an exciton-polariton. In near-planar microcavities, sufficient pump power leads polariton condensation~\cite{Carusotto13}. Condensation is considered distinct from lasing in that the excitons interact with each other substantially (see Ref.~\cite{MicrocavitiesBook}, p362), approaching thermal equilibrium, even if imperfectly. The excitons associated with the condensed polaritons can be free to move (Wannier excitons, typical of inorganic semiconductors~\cite{Kasprzak06}) or bound to individual sites (Frenkel excitons, typical of organic fluorescent solids~\cite{Daskalakis14, Plumhof14}). By contrast, thermalization and BEC of photons as described above is performed in the weak-coupling limit, and with liquid-state matter. 

Classical wave condensation, sometimes known as Rayleigh-Jeans condensation, of light is a nonlinear wave phenomenon. Let us consider light with spatial amplitude or phase noise, propagating through a nonlinear medium. Stochastically, the spatial spectrum will redistribute to follow a Rayleigh-Jeans distribution in transverse wavenumber $k$: \mbox{$f(k) = T / \left( \epsilon_k -\mu \right)$} where $T$ is the amount of noise which is equivalent to a temperature, $\epsilon_k \propto k^2$ the equivalent of kinetic energy, and $\mu$ represents the total light power propagating relative to the nonlinearity. It has been achieved using simple imaging optics~\cite{Sun12} and in multimode pulsed lasers~\cite{Weill10}. There is no mention of quantization of light or matter here, and indeed the distribution is the high-temperature limit of Bose-Einstein distribution ($\epsilon_k \ll T$), \eqnref{eqn:BE distribution}, provided that modes are very closely spaced ($\hbar \rightarrow 0$). Since photon BEC shows an exponential decay of population at high energy, it cannot be classical wave condensation.

Very recently, BEC of plasmons has been achieved, using a lattice of metallic nanoparticles~\cite{Hakala17}, laser pumped and immersed in a bath of fluorescent dye. The band structure of this lattice shows a quadratic dispersion. Thermalization of plasmons occurs via scattering of light from the dye, in very much the same way as for dye-microcavity photon BEC.

\section{State of the art}

Having presented the physical principles and experimental basics, we now review the history and state of the art of Photon BEC.

The thermalization of photons in a dye-filled microcavity was first shown by Martin Weitz's group in Bonn~\cite{Klaers10a}, far below condensation threshold. Proof of thermalization relied on showing that the distribution of light in the cavity is largely independent of the details of the pumping, e.g. pump light position. They also showed that thermalization works well only with detuning of the cutoff wavelength close to resonance so that re-absorption of a cavity photon is likely to happen before loss from the cavity. Having proven thermal equilibrium, they cranked up the pump power. Macroscopic occupation of the ground state, much as in \figref{fig:characteristic results} was observed~\cite{Klaers10b}. Together with the thermal equilibrium, that is sufficient evidence for most commentators to declare that BEC has been achieved~\cite{Anglin10}. In addition, they inferred a thermo-optic repulsive interaction between photons, whose value $\tilde{g}$ in dimensionless units is $\tilde{g} \simeq 7\times 10^{-4}$ which is very small indeed and indicate that photon BEC is for the most part, an ideal gas of non-interacting bosons.

Since these initial observations, there have been a great number of theoretical discussions of how photon BEC happens, and what one expects its properties to be. Measurements have been rarer, with only the Weitz group and ours publishing experimental articles on the topic, with Dries van Oosten's group (Utrecht) having more recently achieved photon BEC.

There have been a small number of review articles on photon BEC, some of which explain the concepts of photon BEC for a non-specialist audience~\cite{Rajan16}. Jan Klaers's tutorials~\cite{Klaers11,Klaers14} provide an excellent introduction to the field. Schmitt \textit{et al.}~\cite{Schmitt16review} is more up to date. There are a few chapters of the book on \textit{Universal Themes of Bose-Einstein Condensation}~\cite{UniversalBECbook} which are relevant to photon BEC and available open-access, most notably Ref.~\cite{Klaers16}. In this section, we attempt a more comprehensive review, covering the majority of the published literature directly on the topic of photon BEC, even if some of the theory work may have little hope of experimental implementation.

\subsection{Observed phenomena}


After their first observation of thermalization and condensation of photons, the Weitz group attempted to move from a liquid to a solid-state sample of dye dissolved in a UV-setting polymer, cured while inside a microcavity~\cite{Schmitt12}. The thermalization functions equally well, although the concentration of dye is limited by fluorescence quenching, as explored in Ref.~\cite{Palatnik16}. While they did observe BEC, it was not reproducible as the dye photobleached at high pump intensity in a matter of seconds. They have also made progress by dissolving materials with large thermo-optic effects to the dye solution, and then locally heating. The resultant position-dependent refractive index translates to a controllable potential energy landscape for the microcavity photons~\cite{Klaers13}.

They then measured both second-order correlations and number fluctuations of the condensate mode~\cite{Schmitt14}. With thermalizing photons it is possible to interpolate between canonical and grand-canonical ensembles by changing the cavity detuning from resonance, effectively changing the ratio of photons to molecular excitations. The molecular excitations form a reservoir. In the canonical ensemble, far detuned from resonance, the photon number is large relative to the square-root of the excitation number, and fluctuations are largely Poissonian. By contrast, close to resonance, the reservoir of excitations is large, and the fluctuations super-Poissonian. The result is that the condensate number can fluctuate wildly, even leading to the Grand Canonical Catastrophe where frequently there are no photons at all in the condensate. While this work was guided by earlier statistical modelling~\cite{Klaers12}, the measured correlations have also been explained through photon-photon interactions~\cite{VanDerWurff14}. The conclusion is that the photon-photon interaction is certainly weak ($\tilde{g} < 10^{-3}$), depends on the detuning from resonance, and that perhaps counter-intuitively the fewer molecules involved, the stronger the interactions.

Two studies indicated how thermalization happens, and how it breaks down. At Imperial College, we produced the first photon condensates outside the Weitz group, and showed how simple parameters such as the shape of the pump spot affect the distribution of photons~\cite{Marelic15}. Light is imperfectly redistributed from pump spot towards the thermal equilibrium distribution. One can achieve condensation with very low pump powers using a small spot, but only for larger spots does the spatial distribution of photons match thermal equilibrium. Using a streak camera and 15-ps pump, Schmitt \textit{et al.}~\cite{Schmitt15} observed the dynamics of thermalization of photons, showing how thermalization happens on the timescale of photon absorption by dye molecules.

BEC is a thermodynamic phase transition. Damm \textit{et al.}~\cite{Damm16} measured the internal energy of the photons (from the spectrum) and defined an equivalent for heat capacity, as a function of not absolute temperature but temperature relative to threshold for condensation. From this, they inferred a heat capacity, which shows a lambda transition characteristic of BEC.

\begin{figure}[htb]
	\centering
	\includegraphics[width=0.75\columnwidth]{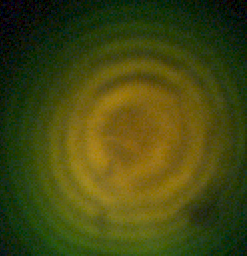}
	\caption{
	Image of the interference pattern of a thermal cloud of photons away from the interferometer white-light fringe. Rings appear because the potential energy landscape is rotationally symmetric, with increasing energy near the edges.
	}
	\label{fig:interferometer image}
\end{figure}

Condensates are typically characterized by their long-range coherence, first hinted at in photon BEC by a single image in Ref.~\cite{Klaers11}. Marelic \textit{et al.}~\cite{Marelic16a} systematically studied stationary first-order coherence using imaging interferometers with slow cameras. They showed that non-dissipative thermal Bose gas theory describes the data well below and just above threshold, with the condensate showing long-range spatial and temporal coherence. Below threshold, the thermal cloud has a position dependent potential energy, which makes for interesting images but complicated analysis: see \figref{fig:interferometer image}. Far above threshold the coherence decreases, which can be explained by multimode condensation, in which several modes become macroscopically occupied. 

Non-stationary measurements of the phase of the condensate mode show phase slips~\cite{Schmitt16}, associated with the occurrences of fluctuations in population number down to zero photons. These phase slips are a clear example of spontaneous phase symmetry breaking in a driven-dissipative system: when the population is large, the phase diffuses only slowly, but with small populations the phase is ill-defined and can jump discontinuously. The condensate re-forms with a spontaneously chosen phase.

Further experiments have explored the momentum distribution of thermalized light~\cite{Marelic16b} showing that the photons interact only weakly with themselves ($\tilde{g} < 10^{-3}$) and with the molecules. More recently, a third group, that of Dries van Oosten in Utrecht, have achieved photon BEC. Their preliminary results show that the condensate is polarized, whereas the thermal cloud around it is unpolarized~\cite{Jagers16}.


\subsection{Theoretical models}

The theory works on photon condensates can loosely be divided into those that assume approximate thermal equilibrium and those that don't. There are works which take a fully quantized approach to fluctuations and there are semiclassical mean-field models, and there are those that apply statistical mechanics.

\subsubsection{A nonequilibrium model of photon condensation}

Foremost among the nonequilibrium models is the Kirton and Keeling model, as first presented in Ref.~\cite{Kirton13}. The model starts from conservative dynamics based on a standard cavity QED model, the Jaynes-Cummings model, with multiple emitters and multiple light modes, and with the addition of a phonon degree of freedom associated with each molecule. Molecular electronic state is coupled to vibrational state via a Huang-Rhys parameter (a continuum generalization of a Franck-Condon factor) since the molecule shape is slightly different between ground and excited electronic states. Drive and dissipation are then included via standard Markovian assumptions. 

The resulting master equation contains terms which include cavity loss, pumping of molecules by an external source, decay of molecules via spontaneous emission out of the cavity, and most crucially both emission of light into the cavity and absorption from the cavity. The latter two terms come with amplitudes which depend on the absorption and emission strength of the dye, and it is these processes which lead to thermalization of light. The master equation, realistically involving millions of molecules and thousands of light modes, is far too unwieldy to solve directly, but the averages of populations can be solved for quite efficiently. The solutions of the rate equation for photon populations show thermalization and condensation matching the Bose-Einstein distribution when the rate at which cavity photons scatter from dye molecules is larger than the cavity loss rate. For larger loss rates, a mode can show threshold behaviour, but it is not necessarily the ground state, indicating lasing rather than BEC.

Kirton and Keeling elaborated further results of their model~\cite{Kirton15}, looking at the dynamics of photon populations after a pulsed pumping event, and evaluating both first- and second-order correlations for individual photon modes. In response to observations of the breakdown of thermalization due to inhomogeneous pumping, in both stationary~\cite{Marelic15} and time-resolved experiments~\cite{Schmitt15}, they modified their model to include spatial distributions of pumping and molecular excitation~\cite{Keeling16}. The results match the salient points of the experimental data. They were able to show that the multimode condensation seen in Ref.~\cite{Marelic16a} was due to imperfect clamping of the molecular excited-state population in regions adjacent to the central condensate light mode which leaves the possibility of positive gain for other modes.

Hesten \textit{et al.} used the Kirton and Keeling model to explore a large parameter space, describing a non-equilibrium phase diagram for dye-microcavity photons~\cite{Hesten17}. The phase diagram proved to be particularly rich, with many possible multimode condensate phases in the crossover between well-thermalized BEC and un-thermalized laser states. In particular, they predict decondensation, where population in a mode decreases with increasing pumping rate, due to mode competition. 

A full master equation approach using just a small number of light modes can be tractable. Kopylov \textit{et al.}~\cite{Kopylov15} have worked with two modes, which is enough modes to draw conclusions about condensation but not about thermalization.

\subsubsection{Quantum field-theory models}

There are several papers treating near-equilibrium as a given, and using quantum field theory techniques such as Schwinger-Keldysh~\cite{deLeeuw13} or quantum Langevin~\cite{Chiocchetta14} techniques to access not only the average behaviour but also fluctuations. These techniques are needed to deal with the fact that photon BEC is driven-dissipative system with both pumping and loss processes (like exciton-polariton condensates), rather than a conservative system (like atomic BEC).

The theory group of Henk Stoof have applied the Schwinger-Keldysh to calculate the effects of drive and dissipation on both temporal~\cite{deLeeuw14b} and spatial~\cite{deLeeuw14a} coherence. They have also shown how interacting photons in a lattice potential behave differently from the superfluid-Mott insulator transition known from conservative systems~\cite{deLeeuw15}. 

The fluctuations of a system at thermal equilibrium are understood to be related to compressibilities and susceptibilities via the temperature in what are known as fluctuation-dissipation relations. Chiochetta \textit{et al.}~\cite{Chiocchetta15} propose testing the fluctuation-relations as a means of quantifying how close driven-dissipative systems like photon BEC come to true thermal equilibrium.

Snoke and Girvin~\cite{Snoke13} point out that it is rather unusual that coherence in a photon BEC can build up despite the absence of direct photon-photon interactions. They show that, remarkably, the coherence is generated through incoherent interactions with the thermal bath of molecular vibrations. 

\subsubsection{Mean-field models}

The equation of motion for the condensate order parameter is typically derived in the same  way as for nonlinear optical systems, and is sometimes known as the Lugiato-Lefever equation~\cite{Lugiato87}, or a dissipative Gross-Pitaevskii equation or a complex Ginzburg-Landau equation. There are various nearly-equivalent forms, which include:
\begin{align}
  -{\rm i}\hbar\partialderiv{\psi}{t} &= \label{Eqn: cGPE} 
  \hspace{-0ex}\left[ V({\bf r})\!-\!\frac{\hbar^2}{2 m}\nabla^2_\perp 
    + g|\psi|^2 + {\rm i}\!\left(\gamma_{net} -\!
      \Gamma|\psi|^2\right) \right]\psi
\end{align}
where $\psi$ is the order parameter, which is the electric field of the condensate mode; $g$ the strength of interactions; $\gamma_{net}$ is the difference between the pump rate and cavity decay rate; $\Gamma$ describes the saturation of molecular excited states; and $m$ and $V$ are effective mass and potential as described earlier. The effective kinetic energy $\nabla^2_\perp$ comes from diffraction of light. The dissipative terms modify beyond-mean-field properties such as correlations and depend on the fact that the pump light is incoherent with the condensate mode~\cite{Carusotto13}. Excluding the dissipative terms, the order parameter equation reduces to \eqnref{eqn:particle energy} for plane waves, ignoring the rest-mass energy term.

Similar equations were first derived for multimode lasers, then applied to light in microcavities~\cite{Chiao99}. Its solutions are Bogoliubov modes of sound~\cite{Chiao00, Bolda01} or collective breathing~\cite{Vyas14} or scissors~\cite{deLeeuw16} modes. It can be derived from Maxwell's equations~\cite{Nyman14}, and coincides with the mean of the equations coming from quantum field treatments~\cite{Chiocchetta14, deLeeuw13}. Interactions in photon BEC are expected to include retarded thermo-optic effects, and nonlocal effects have also been considered~\cite{Strinati14}.

\subsubsection{Statistical models}

It is possible to treat photon BEC with non-quantum formalisms from statistical mechanics or laser rate equations, where quantum effects only come in through bosonic stimulation or exchange statistics. Average populations for effectively two-dimensional~\cite{Kruchkov14} and one-dimensional~\cite{Kruchkov16,Cheng16} landscapes are readily calculated from the Bose-Einstein distribution.

Fluctuations in numbers of photons are correctly calculated only if the finite size of the reservoir of molecular excitations is taken into account~\cite{Klaers12, Sobyanin12}. When taking into account polarization modes of the light, there is at least one prediction that the second-order correlations of condensed light could show sub-Poissonian statistics (anti-bunched) with not unreasonable parameters~\cite{Sobyanin13}. Although the approximation that photons do not interact among themselves is both simplifying and usually applicable to photon BEC, it has been shown that interacting photons should show non-Gaussian statistics, and perhaps suppress the Grand Canonical Catastrophe~\cite{Weiss16, Zannetti15}.

\subsection{Suggestions for alternative systems for photon condensation}

So far, experiments on photon BEC have been restricted to near-planar microcavities filled with one of a small number of fluorescent dyes (mostly Rhodamine 6G and Perylene Red) in liquid water or ethylene glycol, with the exception of Ref~\cite{Schmitt12} in a UV-set polymer. The requirements of the thermalizing medium are rather general: satisfying the Kennard-Stepanov/McCumber relation, having a good fluorescence quantum yield and strong absorption at high concentrations. Other dyes and perhaps colloidal quantum dots are obvious candidate replacement materials. Suitable media may also include optomechanical devices~\cite{Weitz13}. Preliminary measurements suggest that both molecular gases at high pressure with ultraviolet light~\cite{Wahl16} and erbium-doped fibres in the infrared~\cite{Weill16} would be suitable media. BEC of photons thermalizing by scattering from plasmas is probably the oldest of all the proposals~\cite{Zeldovich69} but still relevant~\cite{Mendonca17}.

The optical environment need only provide a minimum energy mode and a gap as well as retaining photons longer than the re-scatter time from the thermalizing medium. For example, planar photonic crystals filled with semiconductors have been proposed for photon thermalization and condensation~\cite{deLeeuw16}, as have arrays of superconducting qubits~\cite{Marcos12}.

There are a few outlandish theoretical proposals to combine photon BEC with quantum optomechanics~\cite{Fani16, Fani17} or atomic BEC~\cite{Zhang16, Zhang12} but, while not technically impossible, it is unlikely that anyone will go to the effort to implement the ideas experimentally.

One unusual proposal~\cite{Chiocchetta16} interpolates between the classical Rayleigh-Jeans condensation of waves and quantum BEC. Light with spatial noise propagates in a non-linear medium with a gradient of linear refractive index perpendicular to the propagation direction, and light is made to selectively leak out for large transverse wavenumbers. This system is then formally equivalent to evaporative cooling of trapped, interacting bosons in two dimensions, where the propagation direction plays the role of time.

\section{When photon BEC gets small}

A question that is often asked about photon BEC is `how is it not a laser'? There are many answers, but it is not unreasonable to argue that photon BEC systems are a very special case of a laser, where the re-scatter of photons is rapid enough to redistribute the light among many cavity modes. But if we look at very tight confinement, i.e. small mirror radii of curvature, the mode spacing can become as large as the thermal energy, and only one cavity mode has significant occupation. In this regime, photons in dye-filled microcavities can exhibit BEC with very small numbers of photons far away from the thermodynamic limit, and they can also act as microlasers, where the spontaneous emission is more likely to go into a cavity mode than in free space. 

In this section we will first see how the concept of threshold in an equilibrium, Bose-Einstein distribution becomes ambiguous for large mode spacings. We will then take a look at a simple rate-equation model which shows microlasers exhibit remarkably similar effects.

\subsection{Tiny Bose-Einstein condensates}

\begin{figure}[htb]
	\centering
	\includegraphics[width=0.95\columnwidth]{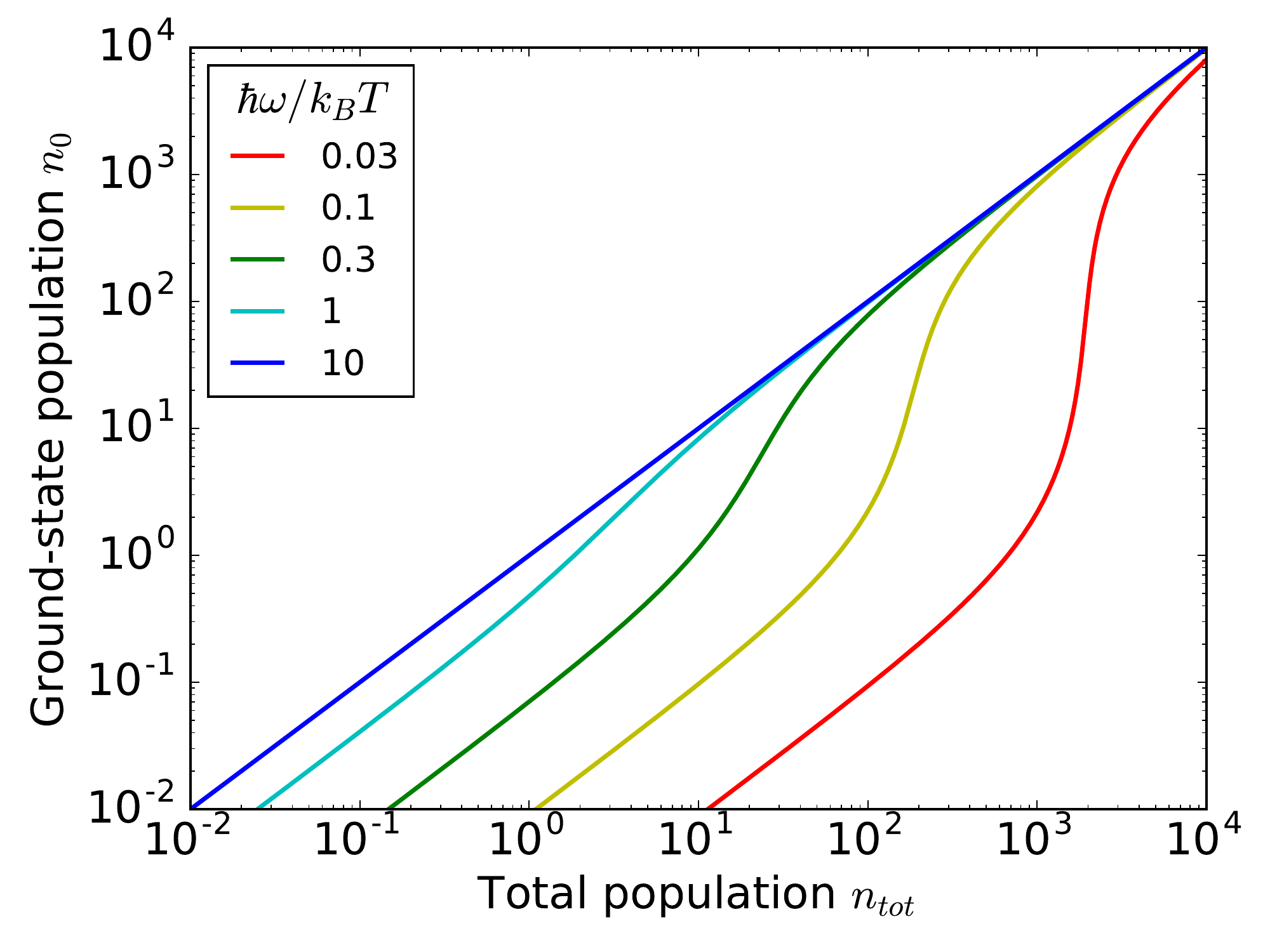}
	\caption{
	The ground-state population in the Bose-Einstein distribution in a two-dimensional harmonic oscillator potential, as a function of total particle number. When the mode spacing is small compared to the temperature, the threshold tends a thermodynamic (sharp) transition. Conversely, for very large mode spacings, only one mode is occupied and no threshold is apparent in the population of the ground state.
	}
	\label{fig:BE distribution}
\end{figure}

The thermal-equilibrium Bose-Einstein distribution \eqnref{eqn:BE distribution} is the very simplest statistical model relevant for photon condensation. In \figref{fig:BE distribution} we show the result for a two-dimensional harmonic oscillator potential of angular frequency $\omega$, where the states involved are discrete with the $i$th state having and energy $i\times \hbar\omega$ and degeneracy $i+1$. We choose a chemical potential, and from that calculate the total population and the ground-state number, as displayed. A well-known result is that the total number of particles in the system at threshold is $N_C = \left({\pi^2}/{6}\right)\left( \hbar\omega / k_B T \right)^{-2}$.

With a small mode spacing (or equivalently a high temperature), the threshold is deep and narrow in the sense that there is a large jump in population for a small change in total population. In the thermodynamic limit, of infinitesimal mode spacing, the threshold is infinitely sharp, and is a true phase transition. On the other hand, for small mode spacing (low temperature), the difference between below- and above-threshold populations is indistinct, and there is a wide range of population where it is not clear if the system is above or below threshold: the threshold is broad and shallow. For extremely small systems, there is just one mode with non-negligible population, so the population of that mode is equal to the total. In that case, there is no threshold in terms of average population, although there may be distinctive correlation or fluctuation behaviour.

\subsection{Microlasers}

Photon BEC takes place inside microcavities, where the spontaneous emission from the dye molecules is modified by the resonator, an effect known as the Purcell effect, which can lead to enhancement (on resonance) or reduction (off resonance) of the spontaneous emission rate~\cite{MicrocavitiesBook}. The factor by which the emission is sped up depends on the cavity parameters (the Purcell Factor, $F_P$) and on the exact position of the molecule in the cavity mode, i.e. the emission rate for a molecule at a node of a cavity mode is very different from that of a molecule at an antinode. The latter factor can vary greatly, so it is difficult to make better than order-of-magnitude estimates for the overall emission enhancement. $F_P$ notably depends inversely on the cavity mode volume: smaller cavities result in large modifications to the emission rate, provided they have large quality factors.

Lasers using microcavities are parameterized principally by the fraction of spontaneous emission directed into the one cavity mode of interest, given the symbol $\beta$. With Purcell enhancement, $\beta = F_P / (1+F_P)$. For large laser systems, where the cavity does not markedly affect the spontaneous emission rate, $F_P\ll 1$ and so $\beta \ll 1$. 

The simplest rate equation model for a microlaser with a single cavity mode containing photon number $P$ interacting with a number of molecular excitations $N$ is:
\begin{align}
  \dot{P} &= [\gamma\beta N - \kappa] P  + \gamma \beta N\label{eqn:microlaser rate photons}\\
  \dot{N} &= R_p - \gamma N - \gamma\beta N P \label{eqn:microlaser rate excitations}
\end{align}
where $\gamma$ is the total spontaneous emission rate including the effects of the cavity, and $R_p$ the pumping rate. Recasting \eqnref{eqn:microlaser rate photons} we find \mbox{$\dot{P} = \gamma \beta N (P+1) - \kappa P$} where the terms in parentheses make clear the roles of stimulated ($P$) and spontaneous ($+1$) emission. This model neglects re-absorption by the fluorescent medium, non-radiative loss, saturation of excited state population and fluctuations but still captures the essential behaviours~\cite{deMartini88, Yokoyama89, Yokoyama91, Yokoyama92, deMartini92, Bjork94, Rice94}.

The equations are readily solved for the steady state population:
\begin{align}
  P = \frac{
	(\beta \rho - 1) + 
	  \sqrt{ (1 - \beta \rho)^2 + 4\beta^2 \rho}
	}{2\beta}
\end{align}
where $\rho = R_p / \kappa$, the rate at which molecules are excited in units of the cavity loss rate. The positive root of the quadratic equation is taken since $P>0$. 

\begin{figure}[htb]
	\centering
	\includegraphics[width=0.95\columnwidth]{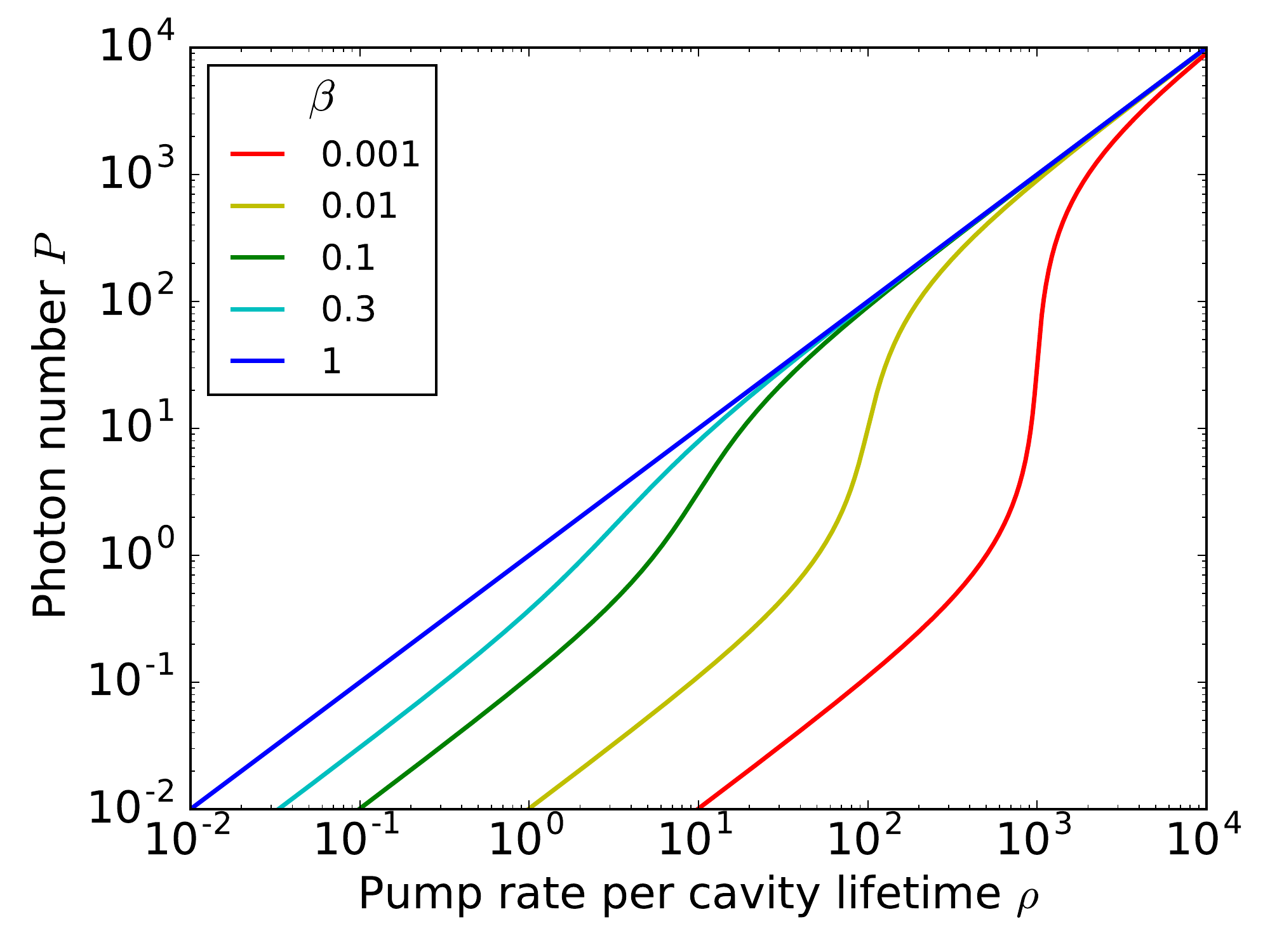}
	\caption{
	Mode population in the microlaser model as a function of pump rate in units of cavity lifetime, for various fractions $\beta$ of spontaneous emission directed into the cavity mode (as opposed to into free space). When $\beta$ is small , the threshold tends a sharp transition. Conversely, for $\beta \rightarrow 1$, all light is directed into the cavity mode and no threshold is apparent in the population.
	}
	\label{fig:microlaser}
\end{figure}
The result is shown in \figref{fig:microlaser}. For comparison, standard lasers have $\beta = 10^{-5}$ -- $10^{-8}$. For small $\beta$, the laser shows a clear threshold, with a large jump in population for a small change in pump rate. When more of the spontaneous emission is directed into the cavity mode, $\beta \rightarrow 1$ and the threshold is less clear, being both shallow, meaning that threshold is accompanied by only a small increase in population, and broad, so over a large range of power it is unclear if the system is above or below threshold.

When all of the spontaneous emission goes into the cavity, there is no obvious threshold. Two competing ideas can be invoked: either this is a thresholdless laser~\cite{Rice94} or we can define the threshold in any reasonable way, for example when the population of the mode exceeds unity~\cite{Bjork94}. In any case, the laser (stimulated emission) action happens at very low pump powers, which is where the industrial interest in microlasers may come from. 

\subsection{Which system is smaller?}

\begin{figure}[htb]
	\centering
	\includegraphics[width=0.95\columnwidth]{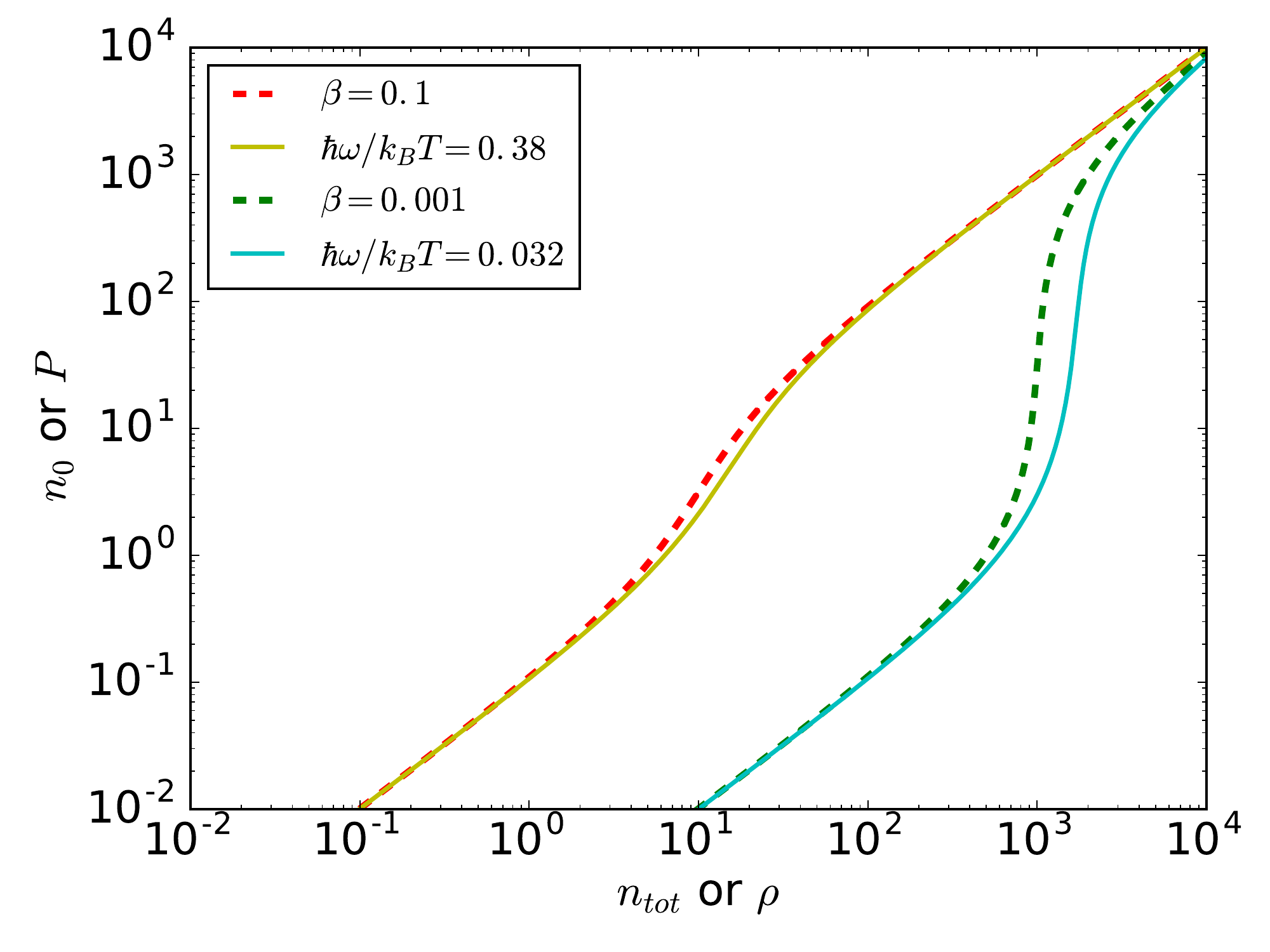}
	\caption{
	Comparison of the two models, BEC and microlaser. $n_{tot}$ and $n$ are the total population and ground-state population (BEC model); $\rho$ and $P$ are pump rate and mode population (microlaser model). For a chosen $\beta$, we set $\hbar\omega/k_B T$ to match the low-number population, and there are no other adjustable parameters.
	}
	\label{fig:comparison}
\end{figure}

Figs.~\ref{fig:BE distribution} and \ref{fig:microlaser} show that tiny Bose-Einstein condensates and microlasers exhibit very similar behaviours in terms of reduction and broadening of threshold when the parameter indicates that the system is size, respectively $k_B T/\hbar \omega$ and $1/\beta$, becomes small. In \figref{fig:comparison} we show results of the models side-by-side. A value of $\beta$ is set, and then $\hbar \omega / k_B T$ adjusted to match mode population in the limit of low pump rate or total population. For small systems, the two models very nearly coincide, although there are deviations for larger parameters. It is therefore difficult to distinguish BEC from lasing, although saturation of excited state populations, as in \figref{fig:characteristic results}, may be a hint. The number of modes thermally available in the BEC model is approximately $(k_B T/\hbar \omega)^2$, and perhaps it is this parameter which should be more directly compared to $1/\beta$. In this respect, BEC is an exclusively multi-mode phenomenon, but if there is only one occupied cavity mode, then there really is not much difference between BEC and lasing.

Microcavities suitable for photon BEC can be constructed using focussed ion beam milling to pattern the confining potential through the surface shape~\cite{Dolan10}. Notably, Ref.~\cite{Palatnik17} shows a solid-state dye microlaser operating in a regime of strong re-absorption, showing features reminiscent of thermalization and BEC. With small radii of curvature, near-single-mode operation, i.e. $k_B T / \hbar \omega \sim 1$, is certainly possible. Very small mode volumes and high quality factors can be simultaneously achieved~\cite{Coles15}. While the bare Purcell factor can be large $F_p\gg 1$, in a fluorescent dye, rapid dephasing due to vibrational relaxation makes it difficult to predict exactly what proportion of the spontaneous emission will be emitted into the cavity mode. Nonetheless, lasers using these microcavities should show $\beta$ parameters which approach unity. It seems that is possible to make a device which can be tuned between tiny BEC and tiny laser, by tuning for example the re-scattering rate via the detuning from the molecular resonance. At threshold, the photon numbers will be rather similar, with barely more than one photon in the lowest-energy cavity mode, despite the different physical origins of the threshold behaviour.

\section{Conclusions}

We have seen how photons can be made to thermalize and condense at room temperature. There is a growing body of literature on this subject, which is connected to wider fields of driven-dissipative condensates of light. When we push the concept of photon BEC to fewer photons, we run into ideas from microlasers, and the distinction between the two concepts becomes blurred, despite the fact that BEC is an equilibrium phenomenon and lasing is dynamic. In this regime of few photons, we expect to find interesting quantum correlations among the photons which may lead to applications of photon BEC in quantum information processing.

\section*{Acknowledgements}

We thank the UK Engineering and Physical Sciences Research Council for supporting this work through fellowship EP/J017027/1 and the Controlled Quantum Dynamics CDT EP/L016524/1 which was co-directed by Danny for many years.

\bibliographystyle{prsty}
\bibliography{photon_bec_refs}

\end{document}